\documentclass[prl,twocolumn]{revtex4}

\usepackage{amsmath,amssymb}
\usepackage{times}

\newcommand{\elem}[2]{\ensuremath{{}^{#2}\text{#1}}}

\begin{document}
\title{Reply to Comment on ``Ab Initio Study of \elem{Ca}{40} with an Importance Truncated No-Core Shell Model''}

\author{R. Roth}
\email{robert.roth@physik.tu-darmstadt.de}
\affiliation{Institut f\"ur Kernphysik, TU Darmstadt, Schlossgartenstr. 9,
64289 Darmstadt, Germany}

\author{P. Navr\'atil}
\email{navratil1@llnl.gov}
\affiliation{Lawrence Livermore National Laboratory, P.O. Box 808, L-414, Livermore, CA 94551, USA}

\maketitle

In their comment on our recent Letter \cite{RoNa07} Dean \emph{et al.} \cite{Dean_comm} criticize the calculations for the ground-state energy of $^{40}$Ca within the importance truncated no-core shell model (NCSM). In particular they address the role of configurations beyond the 3p3h level, which have not been included in the $^{40}$Ca calculations for large $N_{\max}\hbar\Omega$ model spaces. 

Before responding to this point, the following general statements are in order. For the atomic nucleus as a self-bound system, translational invariance is an important symmetry. The only possibility to preserve translational invariance when working with a Slater determinant basis is to use the harmonic oscillator (HO) basis in conjunction with a basis truncation according to the total HO excitation energy, i.e. $N_{\rm max}\hbar\Omega$, as done in the {\it ab initio} NCSM. This is important not only for obtaining proper binding or excitation energies, but also for a correct extraction of physical wavefunctions. The spurious center-of-mass components can be exactly removed only if the HO basis and the $N_{\rm max}\hbar\Omega$ truncation are employed.  
The minimal violation of the translational invariance was one of the main motivations for developing the importance-truncation scheme introduced in the Letter. In this scheme, we start with the complete $N_{\rm max}\hbar\Omega$ HO basis space and select important configurations via perturbation theory. All symmetries are under control and our importance-truncated NCSM calculations are completely variational and provide an upper bound of the ground-state energy of the system. 

The restriction to the 3p3h level, made for computational reasons in the $N_{\rm max}>8$ calculations for $^{40}$Ca, is not inherent to the importance truncation scheme. The explicit inclusion of 4p4h configurations---though computationally more demanding---is straight-forward, even for the largest $N_{\rm max}\hbar\Omega$ model spaces discussed. To demonstrate this fact we have performed full 4p4h calculations for $^{40}$Ca in a $14\hbar\Omega$ no-core model space at $\hbar\Omega=24$~MeV using the $V_{\text{low}k}$ interaction employed in the Letter. The resulting ground state energy of $E_{4p4h}=-471.0$~MeV can be compared with $E_{3p3h}=-461.2$~MeV for the 3p3h calculation reported in Fig. 5(b) with an uncertainty of typically $1$~MeV due to the extrapolation $\kappa_{\rm min}\to0$. Thus the 4p4h configurations change the resulting ground-state energy of $^{40}$Ca by approximately $2\%$. 

In addition to the explicit inclusion of 4p4h or higher-order configurations in the importance truncated space, there are various other means to account for their effect on the energy. We are presently developing a perturbative method for the inclusion of up to 6p6h configurations using the ground state obtained from the NCSM diagonalization as unperturbed state. Furthermore, there are even simpler methods, such as multi-reference Davidson corrections, which are employed successfully in quantum chemistry \cite{JaMe85}, for estimating the contribution of excluded configurations to the energy. They even restore size extensivity in an approximate way. 

The coupled-cluster method (CCM) used by the authors of the Comment lacks the above discussed features important for the nuclear many-body problem, in particular it violates translational invariance from the very beginning and it does not fulfill the variational principle. The problem of a spurious center-of-mass contamination of the many-body states in CCM is not resolved and often not even mentioned (see e.g. Ref. \cite{Hage07}). Moreover, non-iterative triples corrections like CCSD(T), as referred to in the comment, tend to overestimate the correlation energy or even collapse. It is claimed that the CCM is very accurate. A closer inspection of recently published nuclear many-body results does not support this statement. In Ref.~\cite{Dean04}, the $^4$He binding energy with the chiral N$^3$LO NN potential was determined with an uncertainty of several MeV. The same is obtained in the {\it ab initio} NCSM with accuracy of 10 keV \cite{NC04,QN07}. The CCM binding energy results for $^{16}$O with the identical chiral N$^3$LO NN potential obtained in Refs.~\cite{W05} and \cite{G06} differ by more than $5$~MeV. According to Ref.~\cite{G06}, the CD-Bonn NN potential overbinds $^{16}$O. This is contrary to results obtained in Ref.~\cite{F04} and also contrary to expectations as the CD-Bonn underbinds lighter nuclei. 

The CCSD(T) ground state energy for $^{40}$Ca with $V_{\rm lowk}$ reported in Ref. \cite{Hage07} is about $30$~MeV lower than our 4p4h-result (CCSD is about $20$~MeV lower). Taking into account the small 2\% difference of our 3p3h and 4p4h results and the violation of the variational principle and the translational invariance in the CCM method, we believe that the CCM result overestimates the exact ground-state energy in this case. Since the CCM results for $^{40}$Ca were obtained only after our Letter was submitted and posted on the eprint archive we have only compared to the CCM result for $^{16}$O which was available beforehand as a private communication.

In conclusion, the importance truncation provides an efficient way to extend the domain of NCSM calculations to medium-heavy nuclei while preserving translational invariance and the variational principle and allowing for systematic and controlled improvements. 

Supported by the Deutsche Forschungsgemeinschaft through contract SFB 634 and by the Department of Energy under Grant DE-FC02-07ER41457. This work in part performed under the auspices of the U.S. Department of Energy by Lawrence Livermore National Laboratory under Contract DE-AC52-07NA27344.

\end{document}